\documentclass[a4paper,11pt]{article}
\pdfoutput=1 

\usepackage{jheppub} 
\usepackage{graphicx,color}
\usepackage{enumitem}
\usepackage{braket}
\usepackage{slashed}
\usepackage[utf8]{inputenc}
\usepackage{hyperref}
\usepackage{float}
\usepackage{caption}
\usepackage{subfigure}
\usepackage{makecell}
\usepackage{epsfig}
\usepackage{amsmath,amssymb}
\usepackage{hyperref}
\usepackage{diagbox}
\usepackage{enumitem}
\usepackage{listings}

\usepackage{xcolor}

\definecolor{codegreen}{rgb}{0,0.6,0}
\definecolor{codegray}{rgb}{0.5,0.5,0.5}
\definecolor{codepurple}{rgb}{0.58,0,0.82}
\definecolor{backcolour}{rgb}{0.95,0.95,0.92}

\lstdefinestyle{mystyle}{
    backgroundcolor=\color{backcolour},   
    commentstyle=\color{codegreen},
    keywordstyle=\color{magenta},
    numberstyle=\tiny\color{codegray},
    stringstyle=\color{codepurple},
    basicstyle=\ttfamily,
    breakatwhitespace=false,         
    breaklines=true,                 
    captionpos=b,                    
    keepspaces=true,                 
    numbers=left,                    
    numbersep=5pt,                  
    showspaces=false,                
    showstringspaces=false,
    showtabs=false,                  
    tabsize=2
}

\hypersetup{
	colorlinks=true,
	citecolor=black,
	filecolor=black,
	linkcolor=blue,
	urlcolor=blue
}

\usepackage{dsfont}

\newcommand{\ee}{\text{e}}

\usepackage{comment}

\def\bZ{\mathbb{Z}}
\newcommand{\I}{\text{i}}
\newcommand{\kom}{\, ,\quad }
\newcommand*{\raw}{\rightarrow}

\newcommand*{\p}{\mathop{}\!\mathrm \partial}

\newcommand{\bP}{\mathbb{P}}
\newcommand{\bR}{\mathbb{R}}
\newcommand{\cN}{\mathcal{N}}

\newcommand{\cK}{\mathcal{K}}

\newcommand{\cM}{\mathcal{M}}

\newcommand{\ov}{\overline}

\newcommand{\NF}{N_{\text{flux}}}

\usepackage{todonotes}

\title{JAXVacua -- A Framework for Sampling String Vacua}

\author[a]{A. Dubey,}
\author[a,b]{S. Krippendorf,}
\author[c]{A. Schachner}

\affiliation[a]{\footnotesize Arnold Sommerfeld Center for Theoretical Physics, Ludwig-Maximilians Universität, Theresienstr.~37, 80333 München, Germany}
\affiliation[b]{\footnotesize Universitäts-Sternwarte, Fakultät für Physik, Ludwig-Maximilians Universität,
Scheinerstr.~1, 81679 München, Germany}
\affiliation[c]{\footnotesize Department of Physics, Cornell University, Ithaca, NY 14853, USA}

\emailAdd{Abhishek.Dubey@physik.uni-muenchen.de}
\emailAdd{sven.krippendorf@physik.lmu.de}
\emailAdd{as3475@cornell.edu}

\abstract{Moduli stabilisation in string compactifications with many light scalars remains a major blind-spot in the string landscape. In these regimes, analytic methods cease to work for generic choices of UV parameters which is why numerical techniques have to be exploited. In this paper, we implement algorithms based on \texttt{JAX}, heavily utilising automatic differentiation, just-in-time compilation and parallelisation features, to efficiently construct string vacua. This implementation provides a golden opportunity to efficiently analyse large unexplored regions of the string landscape. As a first example, we apply our techniques to the search of Type IIB flux vacua in Calabi-Yau orientifold compactifications. We argue that our methods only scale mildly with the Hodge numbers making exhaustive studies of low energy effective field theories with $\mathcal{O}(100)$ scalar fields feasible. Using small computing resources, we are able to construct $\mathcal{O}(10^6)$ flux vacua per geometry with $h^{1,2}\geq 2$, vastly out-performing previous systematic searches. In particular, we showcase the efficiency of our methods by presenting generic vacua with fluxes below the tadpole constraint set by the orientifold with up to $h^{1,2}=25$ complex structure moduli.

}

\begin{document} 

\begin{flushright}LMU-ASC 19/23
\end{flushright}

\maketitle
\flushbottom

\section{Introduction}
\label{Intro}

The string landscape is believed to contain a large number of vacuum solutions, current estimates ranging between $10^{500}$~\cite{Ashok:2003gk,Denef:2004ze} in type IIB flux compactfications to $10^{272,000}$~\cite{Taylor:2015xtz} in F-theory constructions.\footnote{Similar amplification of vacuum solutions has also been observed in other low-energy theories of string theory~\cite{Lerche:1986cx}.}
Despite these statistical estimates, it is fair to say that our capabilities to efficiently construct string vacua are extremely limited. 
To understand the low-energy physics attainable from string compactifications, a comprehensive numerical approach for analysing string theory vacua is therefore highly desirable. This is the overarching objective of this work. Success in this endeavour would enable an unprecedented look into the nature of consistent theories of quantum gravity and, in particular, their distinguishing features compared to bottom-up Effective Field Theories (EFTs).

More specifically, high-performance methods to search for realistic vacua will allow us to explicitly scrutinise existing constructions of e.g.~string inflation~\cite{Baumann:2014nda, Cicoli:2023opf}, while at the same time explore previously unknown mechanisms in the landscape.
In this context, there are several computational steps which have to be addressed simultaneously:
\begin{enumerate}
\item \emph{Sampling UV data (compactification data):} Zooming in on Calabi-Yau (CY) compactifications, large datasets of compactification spaces have been constructed with the most prominent example being the Kreuzer-Skarke (KS) database of 4-dimensional reflexive polytopes~\cite{Kreuzer:2000xy}. To obtain the associated EFTs, the relevant information about these compact dimensions has to be extracted. This includes the choice of geometry, its topological data like intersection numbers as well as suitable D-brane/O-plane setups satisfying tadpole and anomaly cancellation conditions. This UV data has been constructed on an individual basis in the past (see~\cite{Denef:2004dm,Blumenhagen:2008zz, Cicoli:2011qg,Cicoli:2012vw} for examples in Type IIB string theory), and tools have been developed to enable the study of large numbers of examples~\cite{Demirtas:2022hqf,Jefferson:2022ssj,Moritz:2023jdb}. By now this provides access to `all' models in the KS database which is the first step to systematically analyse their vacuum structures. 
\item \emph{Calculating EFT properties (quantum corrections):}
Consistently identifying perturbative and non-perturbative effects associated with a given UV sample is a prerequisite for proper control over the EFT. For CY compactifications, this includes in particular the scalar potential for the moduli fields. For these setups, several interesting phenomenological implications (most prominently an additional phase of matter domination in the early Universe) have been established in concrete geometric constructions (see~\cite{Cicoli:2023opf} for a recent review). On the methodological side, following over thirty years of development, we now have tools available for computing a wide range of complex structure potentials~\cite{Candelas:1990rm,Hosono:1993qy,Gukov:1999ya,Giddings:2001yu}. In addition, to control other moduli (e.g.~Kähler moduli and open string moduli), further effects have to be considered such as non-perturbative effects via gaugino condensation or D-brane instantons~(see~\cite{Blumenhagen:2009qh} for a review).

Many explicit computational methods for various aspects of the EFT are currently under development. This includes for instance Pfaffian pre-factors in non-perturbative superpotentials (see~\cite{Demirtas:2021nlu,Alexandrov:2022mmy,Gendler:2022qof,Kim:2022uni,Demirtas:2023als} for recent advances) or 1-loop corrections to the Kähler potential in 4D $\cN=1$ compactifications~\cite{Kim:2023cbh,Kim:2023sfs,Kim:2023eut}. Even more importantly, the status of $\alpha^{\prime}$ corrections in $\cN=1$ compactifications is of major concern for the success of Kähler moduli stabilisation procedures like KKLT~\cite{Kachru:2003aw} and, in particular, the LARGE volume scenario~\cite{Balasubramanian:2005zx}, see e.g.~\cite{Garcia-Etxebarria:2012bio,Minasian:2015bxa,Klaewer:2020lfg,Cicoli:2021rub} for progress using F-theory dualities. 

\item \emph{EFT analysis:} For many explicit constructions, analysing the EFTs has been a bottleneck due to a plethora of moduli fields which effectively limits our access to string theory solutions. For example, typical examples in the KS database feature ${\cal O}(100)$ complex structure and/or Kähler moduli fields with the Hodge numbers being bounded by $h^{p,q}\leq 491$. For these systems, analytic approaches are generically impossible except under certain assumptions.\footnote{See~\cite{Demirtas:2021nlu} for SUSY AdS vacua with $h^{1,1}\geq 51$.} In general, one faces two significant bottlenecks: on the one hand, the EFT analysis usually has to be hard-coded on an example by example level. On the other hand, the numerical optimisation appeared to be inefficient~\cite{Martinez-Pedrera:2012teo,Cicoli:2013cha}.

\end{enumerate}

In this work, we focus on developing tools which naturally combine the data products from the first two of the above points and, in light of the third item, address the problem of efficiently sampling string vacua. Specifically, we focus on Type IIB flux vacua which are typically used as starting points for many moduli stabilisation procedures~\cite{Giddings:2001yu,Kachru:2003aw,Balasubramanian:2005zx}. We construct UV input from smooth CY orientifolds where no complex structure or Kähler moduli are projected out. Furthermore, we focus on the large complex structure regime in moduli space where the EFT data is easily derived from string dualities~\cite{Hosono:1993qy,Hosono:1994ax,Cox:2000vi}. Our examples are obtained from the KS database using \texttt{CYTools}~\cite{Demirtas:2022hqf} by taking advantage of recent advances in mirror symmetry~\cite{Demirtas:2023als} and orientifold constructions~\cite{Jefferson:2022ssj,Moritz:2023jdb}.

Our main objective is a self-contained framework for computing EFT properties in the aforementioned compactifications. Using automatic differentiation, we connect elementary inputs such as the prepotential $F$ for the complex structure moduli with the Kähler potential $K$ and superpotential $W$. Subsequently, this allows us -- again using automatic differentiation -- to calculate the scalar potential and its derivatives. In our implementation we make heavy use of the just-in-time compilation and parallelisation features of \texttt{JAX}~\cite{jax2018github} which, as discussed later, accelerates our code in comparison to a simple python implementation by more than two orders of magnitude. We highlight that, although we provide a framework for type IIB flux compactifications, all components of our numerical implementation are modular. Little effort is required to build extensions including further aspects of low energy EFTs or exploring different string duality frames.

In our pipeline, we are able to generate efficiently the relevant equations (like the $F$-flatness conditions $D_{I}W=0$) associated with string solutions for any UV input. This defines an underlying optimisation problem which is the main use case in this work. We solve the latter using standard optimisation methods readily available via standard libraries. Arguably, solving this optimisation problem and generating solutions numerically is a challenging task by itself, but it becomes even more difficult to locate \emph{trustworthy} and \emph{phenomenologically relevant} solutions. Regarding the former, this means that we seek to find solutions in regimes of moduli space where calculations are under control. Equally important, we constrain our search to those values of the tadpole that can be realised in actual orientifold compactifications of Type IIB superstring theory. If we allowed unconstrained tadpoles, we would naturally find solutions almost anywhere for any number of moduli simply because of the drastically increased density of vacua in our search space.

We stress that, apart from the minimising task itself, evaluating and checking the aforementioned constraints on, say, billions of fluxes and the associated solutions to F-term equations is computationally expensive. The study of such large ensembles is necessary to address phenomenological requirements and to be able to statistically test the properties of the string landscape in a meaningful manner. Our framework naturally connects with other ongoing work to develop methods which allow to efficiently search for phenomenologically viable solutions in the string flux landscape~\cite{Cole:2019enn,Krippendorf:2021uxu,Cole:2021nnt,Krippendorf:2022gcl}.

The rest of this paper is organised as follows. In Section~\ref{sec:FC} we give a short portrait of CY orientifold flux compactifications and introduce the type of optimisation problems to be solved in the remainder of the paper. Section~\ref{sec:algorithm} describes our implementation and numerical approach to finding flux vacua. We present detailed numerical experiments in Section~\ref{sec:numericalexperiments} and summarise our results in the conclusions in Section~\ref{sec:Con}.

\section{Type IIB flux compactifications}\label{sec:FC}

In this section, we review Type IIB flux compactifications on CY orientifolds and the type of equations that need to be solved for flux vacua. We mainly state results to set our notation and refer to the literature for more detailed discussions (see~\cite{Grana:2005jc,Douglas:2006es,Hebecker:2021egx} for reviews). These considerations are prerequisites for the subsequent implementation of our root finding and sampling methods.

\subsection{Type IIB flux vacua}

Compactifying Type IIB superstring theory on a CY orientifold leads to 4D $\cN=1$ supergravity with $h^{1,1}_{+}$ Kähler and $h^{1,2}_{-}$ complex structure moduli $Z^{i}$.
The classical Kähler potential $K$ for the $Z^{i}$ and the axio-dilaton $\tau$ can be written as
\begin{equation}\label{eq:TreeLevKP}
 K=-\log( -\I\, \Pi^{\dagger}\cdot\Sigma\cdot \Pi ) - \log\left(-\I({\tau}-\ov{\tau})\right) \kom \Sigma=\left (\begin{array}{cc}
0 & \mathds{1} \\ [-0.2em]
-\mathds{1} & 0
\end{array} \right )\, .
\end{equation}
Here,
we defined the period vector $\Pi$ in terms of the prepotential $F$ as
\begin{equation}\label{eq:PeriodVecGen} 
\Pi=\left (\begin{array}{c}
 2F-Z^{i} F_{i}\\ [-0.15em]
F_{i}\\ [-0.15em]
1 \\ [-0.15em]
Z^{i}
\end{array} \right ) \kom F_i = \p_{Z^{i}}F\, .
\end{equation}
The presence of 3-form fluxes induces the $F$-term scalar potential
\begin{equation}\label{eq:FC:33}
V_{\text{Flux}}= \ee^{K} \left ( K^{\tau\bar{\tau}}D_{\tau} W\, D_{\bar{\tau}}\overline{ W}+ K^{i\bar\jmath}D_{i} W\, D_{\bar\jmath}\overline{ W}\right )
\end{equation}
in terms of the Gukov-Vafa-Witten (GVW) superpotential~\cite{Gukov:1999ya}
\begin{equation}\label{eq:SupPotPerVec} 
 W=\left (f-\tau h\right )^{T}\cdot \Sigma\cdot   \Pi(Z) \, ,
\end{equation}
and integer flux quanta
\begin{equation}\label{eq:FC:18B}
f=(m_{J}, m^{I})\kom h=(n_{J},n^{I})\kom I,J=0,\ldots ,h^{1,2}_{-}(X_{3})\, .
\end{equation}
For given choices of a 3-form flux background $G_{3}=f-\tau h$, the potential \eqref{eq:FC:33} exhibits a non-trivial structure of minima called \emph{flux vacua}. In the first instance it is these flux vacua which we aim at identifying. The available choices of fluxes are limited because they contribute to the cancellation condition for the $ C_{4}$-tadpole,
\begin{equation}\label{eq:FC:36}
\dfrac{1}{2}N_{\text{flux}}+N_{D3}-N_{\overline{D3}}=\dfrac{Q_{\text{D3}}}{2}\kom Q_{\text{D3}}=\dfrac{\chi(Y_4)}{12}~,
\end{equation}
in terms of the D3-brane charge induced by fluxes
\begin{align}
  N_{\text{flux}} =\int_{X_{3}}\,  H_{3}\wedge F_{3}= f\cdot\Sigma\cdot h=m_{I}n^{I}-m^{I}n_{I}\, .
\end{align}
where $N_{D3}$ ($N_{\overline{D3}}$) denotes the number of (anti-)D3-branes.
The quantity $\chi(Y_{4})$ is the Euler characteristic of the $F$-theory fourfold~\cite{Denef:2008wq} encoding D3-charge contributions from O-planes and D7-branes.

In this work, we focus on a class of flux vacua for which the $F$-term conditions $D_{\Phi^{I}}W=D_{I} W=0$, $\Phi^{I}\in \lbrace \tau, Z^{i}\rbrace$, are satisfied.\footnote{The study of general vacua (e.g.~non-supersymmetric solutions with $D_{I} W\neq 0$) is left for the future.}
One finds that
\begin{align}\label{eqn:fflat1}
  D_{\tau}  W&=\frac{1}{\overline{\tau}-\tau}(f-\overline{\tau}h)^{T}\cdot \Sigma\cdot \Pi(Z)=0\, ,\\
  D_i  W&= (f-\tau h)^{T}\cdot \Sigma\cdot\left(\partial_{i} \Pi(Z)+\Pi(Z)\partial_{i} K\right)=0\, . \label{eqn:fflat2}
\end{align} 
In fact, these two conditions are equivalent to saying that the $3$-form $G_3$ must be imaginary self-dual (ISD), i.e., $ \star_{6}  G_{3}=\I  G_{3}$ in terms of the Hodge star operator $\star_3$ on $X_3$. We can express this condition more explicitly as~\cite{Grimm:2021ckh}
\begin{equation}\label{eq:ISD_cond_flux} 
m_{J}-\tau n_{J} =\overline{\cN}_{JI}\left (m^{I}-\tau n^{I}\right ) 
\end{equation}
in terms of the gauge kinetic matrix
\begin{equation}\label{eq:GaugeKinMatrix}
\cN_{I J}=\overline{F}_{IJ}+2\I\, \dfrac{\text{Im}(F_{I L})X^{L} \, \text{Im}(F_{J K})X^{K}}{X^{M}\text{Im}(F_{MN})X^{N}}\kom F_{IJ}=\p_{X^{I}}\p_{X^{J}}F\, .
\end{equation}
For ISD fluxes, $N_{\rm flux}\geq 0$ is non-negative which is why sources with negative D3-brane charge like D-branes and O-planes~\cite{Giddings:2001yu} are required for tadpole cancellation \eqref{eq:FC:36}.

Flux choices can lead to equivalent vacua by being identified under transformations of $\mathcal{G}=\text{SL}(2,\mathbb{Z})\times \Gamma$ where the former is the Type IIB S-duality group and $\Gamma$ the modular group acting on $\cM_{\text{cs}}(X_{3})$. This can be avoided by mapping solutions to the fundamental domain of $\text{SL}(2,\mathbb{Z})$. For later convenience, let us define the VEV of the superpotential as\footnote{The convention for the normalisation is chosen based on~\cite{Kachru:2019dvo,Demirtas:2019sip}.}
\begin{equation}
    W_0 = \sqrt{\dfrac{2}{\pi}} \, \biggl < \ee^{K/2}\, W \biggl >~,
\end{equation}
which is manifestly invariant under $\mathrm{SL}(2,\mathbb{Z})$ transformations.

We note that for models with $h^{1,2}>1$, very little is known about the solution space of string theory, see however~\cite{Cicoli:2013cha,Brodie:2015kza,Marsh:2015zoa} for examples with $h^{1,2}\geq 4$ and~\cite{Martinez-Pedrera:2012teo} for $h^{1,2}=2$. A key obstacle hereby is systematically solving $F$-term conditions \eqref{eqn:fflat1}, \eqref{eqn:fflat2} for generic choices of fluxes below tadpole. Typically, analytic methods cease to work in regimes with $h^{1,2}>1$ unless for special classes of vacua like in~\cite{Demirtas:2019sip,Blanco-Pillado:2020hbw,Marchesano:2021gyv,Coudarchet:2022fcl} where a subset of VEVs can be fixed analytically under certain conditions.

\subsection{Explicit models at large complex structure}

In explicit examples, we need to compute the period vector $\Pi$ entering \eqref{eqn:fflat1}, \eqref{eqn:fflat2}. These periods can be computed by solving Picard-Fuchs equations~(cf.~\cite{Hosono:1993qy,Hosono:1994ax,Cox:2000vi}), employing localisation techniques~(cf.~\cite{Jockers:2012dk}) or using asymptotic Hodge theory~(cf.~\cite{Bastian:2021eom}).

An important class of models concerns \emph{Large Complex Structure} (LCS) limits in complex structure moduli space $\cM_{\text{cs}}(X_{3})$. For such scenarios, the analytic structure for the prepotential is well understood and easily computed using mirror symmetry, see~\cite{Demirtas:2023als} for recent progress. In fact, around LCS points, the coordinates $Z^{i}$ of $\cM_{\text{cs}}(X_{3})$ are identified with the Type IIA Kähler moduli in the large volume limit of Type IIA string theory compactified on the mirror dual CY manifold $\tilde{X}_{3}$~\cite{Morrison:1991cd,Hosono:1994av,Hosono:1994ax}. This identification implies that the complex structure moduli $Z^{i}$ take values in the Kähler cone of $\tilde{X}_{3}$
\begin{equation}\label{eq:DefKahlerCone}
    \cK_{\tilde{X}_3} = \lbrace J\in H^{1,1}(\tilde{X}_3,\bR):\, \text{Vol}_{J}(U)>0\; \forall \text{ sub-varieties }U\rbrace \subset H^{1,1}(\tilde{X}_3,\bR)\, ,
\end{equation}
where the sub-varieties are effective curves, effective divisors and $\tilde{X}_3$ itself.
It describes the moduli space of Kähler structures on $\tilde{X}_3$ parametrised by a Kähler form $J$.
By abuse of terminology, we simply refer to $\cK_{\tilde{X}_3}$ as the Kähler cone without mentioning the mirror side explicitly.

The prepotential computed from mirror symmetry reads~\cite{Morrison:1991cd,Hosono:1994av,Hosono:1994ax}
\begin{equation}
\label{eq:prepotentialNew}
F= -\dfrac{1}{6}\kappa_{ijk} \,  Z^i\,  Z^j \,  Z^k +  \frac{1}{2} \,{a_{ij} \,  Z^i\,  Z^j} +  \,{b_{i} \,   Z^i} +  \frac{\I}{2} \, \,{\tilde \xi} +    F_{\text{inst}}(Z)\,.
\end{equation}
Here, the parameters $\kappa_{ijk}$ are the triple intersection numbers of $\tilde{X}_3$ which, along with the other parameters, are defined as
\begin{align}
\kappa_{ijk} &= \int_{\tilde{X}_3} \, J_i \wedge J_j \wedge J_k\kom a_{ij} = \frac{1}{2}\int_{\tilde{X}_3} \, J_i \wedge J_j \wedge J_j\,\text{mod}\,\mathbb{Z}\; ,\quad \nonumber\\
b_j &= \frac{1}{4!}\int_{\tilde{X}_3} \,c_2(\tilde{X}_3) \wedge J_j\kom \tilde{\xi} = \frac{\zeta(3)\, \chi(\tilde{X_3})}{(2\pi)^3}\,.
\end{align}
Here, $c_{2}(\tilde{X}_{3})$ denotes the second Chern class of the mirror manifold $\tilde{X}_{3}$. Further, the $J_{i}\in H^{1,1}(\tilde{X}_{3},\bZ)$ are $(1,1)$-forms and $\chi(\tilde{X}_{3})$ is the Euler characteristic of $\tilde{X}_{3}$.

Finally, the worldsheet instantons on the mirror dual side give rise to exponential contributions of the form~\cite{Hosono:1994av,Hosono:1994ax}
\begin{equation}\label{eq:InstCorrections} 
    F_{\text{inst}}(Z^{i})=-\dfrac{1}{(2\pi \I)^{3}}\sum_{q \in \cM(\tilde{X}_{3})}\, n_{q}^{(0)}\, \text{Li}_{3}\left (\ee^{2\pi \I\, q_i\, Z^i}\right )\kom \text{Li}_{3}(x)=\sum_{m=1}^{\infty}\, \dfrac{x^{m}}{m^{3}}
\end{equation}
in terms of genus zero \emph{Gopakumar-Vafa (GV) invariants} $n_{q}^{(0)}$~\cite{Gopakumar:1998ii,Gopakumar:1998jq} of effective curves $q$ in the \emph{Mori cone} $\cM(\tilde{X}_{3})$ of the mirror manifold $\tilde{X}_{3}$. It turns out to be more convenient to work with a different set of invariants obtained from a resummation of poly-logarithms. These are the so-called (genus zero) \emph{Gromov-Witten (GW) invariants} $N_{q}^{(0)}$ which are related to the GV invariants via
\begin{equation}
    \sum_{q \in \cM(\tilde{X}_{3})}\, n^{(0)}_{q}\, \text{Li}_{3}\left (\ee^{2\pi \I\, q_i\, Z^i}\right )=\sum_{q \in \cM(\tilde{X}_{3})}\, N_{q}^{(0)}\,  \ee^{2\pi \I\, q_i\, Z^i}\, .
\end{equation}
A systematic approach to computing these invariants has been established by HKTY~\cite{Hosono:1993qy,Hosono:1994ax}. In practice, we compute the GV and GW invariants using \texttt{CYTools}~\cite{Demirtas:2022hqf,Demirtas:2023als} up to some finite degree.

The validity of the ansatz \eqref{eq:prepotentialNew} for $F$ is limited to the region of convergence of the LCS expansion~\cite{Hosono:1994av}. The radius of convergence is determined by the singularity of the associated Yukawa couplings~\cite{Candelas:1994hw,Klemm:1999gm}. In this paper, we only check mild conditions to ensure the validity of our solutions up to a given cutoff on the GV invariants. That is, we look at solutions for which
\begin{equation}\label{eq:InstCutoff}
    \biggl |\dfrac{F_{\text{inst}}}{F_{\text{pert}}}\biggl |<\varepsilon
\end{equation}
for some $\varepsilon$ where $F_{\text{inst}}$ is computed up to some finite cutoff in the instanton expansion. In practice, we typically choose $\varepsilon=0.1$, though we also investigate the cutoff dependence of the instanton expansion by including higher order contributions to $F_{\rm inst}$ in the region of interest.

Before we continue, let us stress that going beyond the LCS regime is possible and should be part of future investigations. Here we limit ourselves to the LCS regime as computational tools from mirror symmetry are readily available in this regime~\cite{Demirtas:2023als}. An extension of these tools to other regimes is a clear direction for future developments (see~\cite{Demirtas:2020ffz,Alvarez-Garcia:2020pxd} for work along those lines).

\subsection{Model construction and data-sets}

Having presented the framework for our optimisation problem, we briefly comment on the data-sets which will be studied subsequently. 

\subsubsection*{CY and orientifold construction}

Foremost, we focus on the Kreuzer-Skarke (KS) database~\cite{Kreuzer:2000xy} comprised of 473,800,776 reflexive polytopes in four dimensions. Any fine, regular, star triangulation (FRST) of these polytopes leads to a CY manifold embedded as the anti-canonical hypersurface~\cite{Batyrev:1993oya}. Given that a single polytope can have many FRSTs, it typically features several non-isomorphic CY manifolds, though the actual number of topologically inequivalent ones remains opaque.\footnote{There exist rigorous upper bounds on the number of homotopically inequivalent CYs from polytopes in the KS database~\cite{Demirtas:2020dbm} which have been obtained by bounding the number of inequivalent, fine 2-face triangulations and applying Wall's theorem~\cite{Wall1966}.}

Orientifolds are constructed from $\mathbb{Z}_2$-involutions of Calabi-Yau manifolds. In the context of the KS database, the fixed point loci can often be obtained from simple conditions on the polytope~\cite{Jefferson:2022ssj, Moritz:2023jdb}. Although such involutions are inherited from the ambient toric variety covering only a subset of all involutions, they allow for the construction of a sizable number of orientifolds. We note that, in some cases, orientifold involutions lead to singularities which have to be resolved appropriately (see~\cite{Candelas:1989js,Carta:2020ohw} for detailed discussions). For the examples discussed in this paper, we checked that our orientifolds do not give rise to any such singularities.

For convenience, we consider orientifolds with $h^{1,1}_-=h^{1,2}_+=0$ for which the D7-tadpole is cancelled locally by putting D7-branes on top of the O7-planes. The D3-charge from O-planes and non-Higgsable $\mathrm{SO}(8)$-stacks is given by
\begin{equation}\label{eq:MaxD3Charge}
    Q_{\text{D3}}= 2+h^{1,1}+h^{1,2}\, .
\end{equation}
Throughout the main text, we constructed suitable models from the KS database using \texttt{CYTools}~\cite{Demirtas:2022hqf} making sure that there exists an orientifold with $h^{1,2}_+=0$.\footnote{We thank Jakob Moritz for explaining us the necessary conditions on polytopes allowing for orientifolds with $h^{1,2}_+=0$.} Ultimately, this requirement ensures that the orientifold intersects the LCS patch allowing us to directly apply the formulas stated in previous section.

\subsubsection*{Flux vacua at LCS}

In this work, we are concerned with analysing large samples of vacua to investigate their distributions and attainable properties. It is clear that, due to the absence of generic analytic solutions, this requires dedicated numerical tools. On the aforementioned geometries, the tadpole (through \eqref{eq:MaxD3Charge}) and hence the possible flux vacua are bounded~\cite{Denef:2004ze,Grimm:2021vpn}. This number is generally speaking huge and estimates obtained from continuous flux approximations suggest that the number of flux vacua with $D_{I}W=0$ and $N_{\text{flux}} \leq Q_{D3}$ is given by~\cite{Denef:2004ze}
\begin{equation}\label{eq:DenefDouglasNumVacua}
    \mathcal{N}\bigl (N_{\text{flux}} \leq Q_{D3}\bigl )=\dfrac{(\pi Q_{D3})^{b_{3}}}{b_{3}!\, \pi^{b_{3}/2}}  \int_{\mathcal{M}_{\text{cs}}}\, \det (\mathcal{R}+\omega)\, .
\end{equation}
Here, we defined the 3rd Betti number $b_{3} = 2(h^{1,2}+1)$, the curvature 2-form $\mathcal{R}$ and the Kähler form $\omega$ on moduli space. We refer to~\cite{Conlon:2004ds} for an explicit evaluation of this formula for examples with few moduli.

We stress that the absence of solutions for a given choice of fluxes in our subsequent routine has to be seen in conjunction with the restriction to the LCS regime and to a single Kähler cone phase associated with a single FRST of a polytope. To give some heuristic motivation, let us assume that for each of the $N$ moduli half of the parameter range falls within the strict LCS regime $\text{Im}(Z^{i})>0$. This implies that at large $N$ a uniformly sampled point in moduli space has a probability of $1/2^N$ to fall in the LCS regime. In reality, the angles between two hyperplanes of the cone becomes smaller as we scale up $h^{1,2}$ making the cone even narrower than our previous naive estimate~\cite{Demirtas:2018akl}. 

On top of that, even if points can be efficiently sampled from $\mathcal{K}_{\tilde{X}_3}$, the actual question to ask is: given a choice of fluxes, what are the chances of finding a minimum within the LCS patch traced out by $\mathcal{K}_{\tilde{X}_3}$? In our case, we search for flux vacua in a single toric phase defined by a single FRST of a polytope. Generally, this geometric phase only covers a small subspace of moduli space where the existence of solutions to \eqref{eqn:fflat1}, \eqref{eqn:fflat2} for arbitrary flux choices with $N_{\rm flux}\leq Q_{D3}$ cannot be guaranteed.\footnote{We refer to~\cite{Gendler:2022ztv} for related discussions in the context of 5D black hole solutions in M-theory.} As we discuss later in Sect.~\ref{sec:Largeh12}, this seriously affects our ability to sample vacua at large values of $h^{1,2}$.

\section{Algorithmic approach for string vacua}
\label{sec:algorithm}

Below we outline the scope and functionalities of our current numerical approach. Our code is modular such that each component can be developed further on an individual basis and allows to be integrated into the pipeline to extract physical properties. In short, the rationale is as follows:
\begin{enumerate}[label = (\roman*)]
    \item \emph{Model construction:} We start with a high level input which can either be a custom example or directly interface with existing databases such as the KS database or the CICY database~\cite{Candelas:1987kf,Anderson:2017aux}.
    \item \emph{EFT module:} We calculate relevant quantities in the 4D supergravity EFT starting from a prepotential. Here, this includes the flux superpotential \eqref{eq:SupPotPerVec} and the Kähler potential \eqref{eq:TreeLevKP} for complex structure moduli. Subsequently, we use the known analytical framework to calculate quantities like the $F$-term scalar potential \eqref{eq:FC:33}. This pipeline is implemented such that we can ensure scalability in the presence of many moduli. To do this, we utilise \texttt{JAX} and its just-in-time (\texttt{jit}) and vectorisation (\texttt{vmap}) features. As detailed below, compared to alternative implementations in \texttt{python} or \texttt{Mathematica}, simply evaluating relevant quantities with our methods is faster by several orders of magnitude, see Fig.~\ref{fig:timing_jit}.
    \item \emph{Optimisation module:} We formulate the optimisation conditions for identifying minima of moduli potentials and pass them to optimisation algorithms. A priori there is no optimisation algorithm which is singled out. At this stage, we restrict ourselves to the use of \texttt{scipy.optimize.root} which we find to be rather efficient for our purposes.
    \item \emph{Sampling module:} Of particular importance for the efficiency of our optimisation algorithm is the choice of initial guesses for the moduli VEVs, i.e.,~from where the algorithm starts searching for minima. We present novel methods to choose these points based on the choice of flux quanta using the ISD condition \eqref{eq:ISD_cond_flux}.
    \item \emph{Filter module:} The minima are identified using a certain numerical tolerance of the respective conditions. To warrant that we are dealing with actual minima, we perform additional checks on these candidates as outlined below. Among others, this involves checking the positivity of the Hessian of the scalar potential.
\end{enumerate}
Below, we describe in more detail each of the individual modules (ii)-(v) building our framework for the search for string vacua. (The model construction (i) is largely outsourced using tools readily available in \texttt{CYTools}.) Again we stress that each of these components could be modified without interfering with the other parts of the code. Given our implementation, our current bottleneck when extrapolating to large $h^{1,2}$ is the sampling and root finding which we comment on in due course.

\subsection{EFT module -- moduli potentials in \texttt{JAX}}
\label{sec:jax}

For our implementation, we opted for \texttt{JAX}~\cite{jax2018github} which at its heart employs automatic differentiation to compute gradients of functions. It recognises primitive operations inside functions and applies standard rules for differentiation in such a way to numerically evaluate derivatives rather efficiently. In particular, it avoids finite differences to compute approximate derivatives in numerical differentiation, and it bypasses inefficient expressions in symbolic differentiation which would appear for instance when using \texttt{sympy}.

Overall, this makes the implementation differentiable, i.e.,~derivatives of quantities with respect to intermediate or input variables can be taken and utilised at machine precision. This is ideal for our quest for finding string vacua in 4D supergravity where only a handful of functions are prerequisites, in our case the prepotential \eqref{eq:prepotentialNew}, in order to compute quantities like period vectors, scalar potentials or the Hessian by taking suitable derivatives. In particular, this implies that our code only needs the absolute minimum of functions to be hard-coded, thereby making our implementation extremely versatile. Among others, it is readily applicable to any moduli space limit away from LCS such as conifold regions and, with minor adjustments, can be easily extended to include Kähler moduli.

Apart from the usefulness of auto-differentiation, there are two powerful tools to speed up the implementations.
The first is \texttt{jit}-compilation (standing for \texttt{j}ust-\texttt{i}n-\texttt{t}ime compilation) which, in short, transforms \texttt{python}-functions into language representations via so-called \emph{tracer objects} that record all the operations being performed. This reconstruction improves the execution by sending full sequences of operations to tensorflow's XLA (Accelerated Linear Algebra) compiler~\cite{tensorflow2015-whitepaper}, while in standard \texttt{python}-code each operation is sent to the compiler one at a time.\footnote{We note however \texttt{jax.jit} cannot be applied to any arbitrary function, see the official \href{https://jax.readthedocs.io/en/latest/jax-101/02-jitting.html}{\texttt{JAX} documentation}. For instance, whenever specific values are tested as conditions (e.g., via \texttt{if}-statements), the \texttt{jit}-compilation is not directly applicable but generally workarounds can be identified.} For our purposes, \texttt{jit} ensures that there is only a modest scaling with respect to the number of moduli fields through essentially C++ speed of the code. Additionally, the usage of \texttt{JAX} makes most modules readily usable on the GPU which can further improve the efficiency.

Secondly, while vectorisation can always be implemented manually (by rearranging indices and axes), this can become increasingly tedious when the range or the number of indices changes dynamically. This is where the automatic vectorisation feature \texttt{vmap} of \texttt{JAX} is particularly useful. For example, we may want to evaluate the $F$-term conditions for a single flux at many points or, instead, for several fluxes, but for each flux at a different collection of points. Both cases can be addressed by calling \texttt{vmap} on the various inputs (see the \href{https://jax.readthedocs.io/en/latest/jax-101/03-vectorization.html}{\texttt{JAX} documentation} for examples). Clearly, the upshot of vectorising expressions in our implementation are enormous gains in efficiency as we now illustrate.

\begin{figure}[t!]
    \centering
    \includegraphics[width=0.8\textwidth]{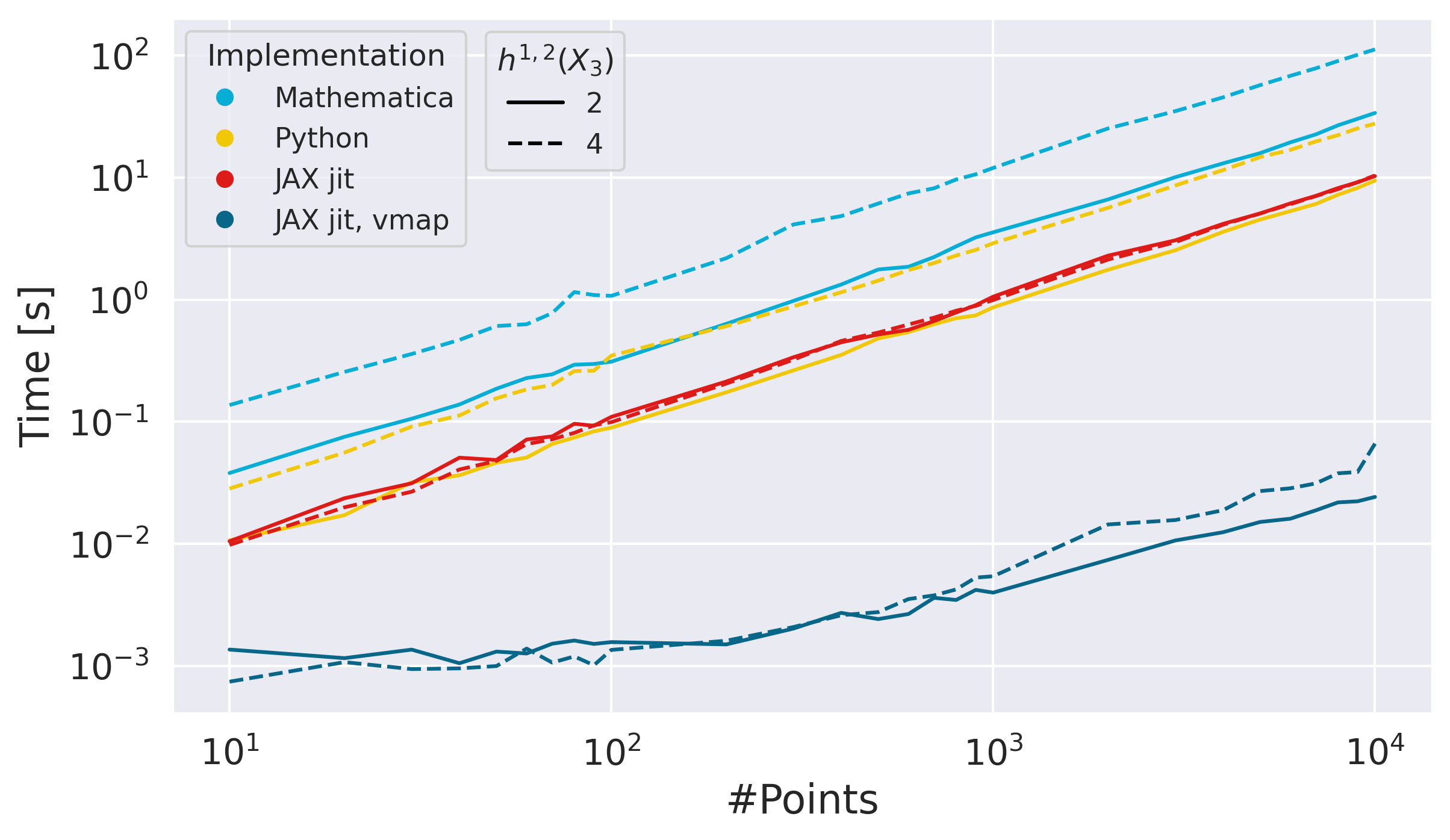}
    \caption{{\bf Efficiency comparison of different implementations:} Time required to evaluate the $F$-term conditions for two models, described in the text, with $h^{1,2}=2,4$ at a given number of points using different implementations. Each curve is obtained from averaging over 10 runs performed on the same laptop using a single core. Further improvements are easily obtained when parallelising over multiple cores or using our code on the GPU.}
    \label{fig:timing_jit}
\end{figure}

Let us quantify the improvements of our implementation compared to popular alternatives, namely \texttt{Mathematica} (version 13.4) and \texttt{python} (version 3.9). We compare the time required to evaluate the $F$-term conditions $D_I W$ for all fields in different implementations in Fig.~\ref{fig:timing_jit}. We present two models with $h^{1,2}=2,4$ complex structure moduli using in both cases GV invariants up to degree 4.\footnote{The two models in question are $\mathbb{P}[1,1,1,6,9]$ for $h^{1,2}=2$ discussed in Sect.~\ref{sec:CP11169} and the first example in~\cite{Cicoli:2013cha} for $h^{1,2}=4$.} In \texttt{Mathematica} and \texttt{python}, we used a hard-coded version of the $F$-term conditions, while our \texttt{jit}- and \texttt{vmap}-compilation starts from the prepotential \eqref{eq:prepotentialNew} and constructs the $F$-terms via automatic differentiation. We use \texttt{JAX} version 0.4.4 in our benchmark comparison.

In both the \texttt{Mathematica} and \texttt{python} implementations, one observes a clear scaling with $h^{1,2}$ coming from the increasing number of primitive operations required when computing $D_IW$. In contrast, the implementations with \texttt{jit} show virtually no difference between the two models. As promised above, the most dramatic speed up is however observed by making use of \texttt{vmap} which improves the evaluation time by roughly two orders of magnitude. A comparison with other implementations like a direct \texttt{C++} implementation is beyond the scope of this paper.

Before we continue, we note that, while we provide implementations for general supergravity equations, it can occasionally be more efficient to use simplified expressions. For example, say we solved the $F$-flatness conditions for a given choice of fluxes. Then it can be useful to employ identities for the $F$-term scalar potential \eqref{eq:FC:33} and related quantities like the Hessian, see e.g.~\cite{deAlwis:2013jaa}.

\subsection{Optimisation module -- numerical search for extremal points}
\label{sec:optimisation}

In our quest for string vacua, we would like to numerically solve $D_{I}W=0$ for given inputs of UV data. Since most optimisation algorithms work with real-valued variables we re-formulate our complex-valued optimisation conditions (cf.~Eqs.~\eqref{eqn:fflat1}, \eqref{eqn:fflat2}) in terms of their respective real and imaginary parts. We end up with a one-dimensional array of size $2(h^{1,2}+1)$ for which we find the roots in terms of the real and imaginary parts of the moduli fields.\footnote{We note that empirically, we find that using the analytic solution for the dilaton obtained from solving \eqref{eqn:fflat1} does not simplify the optimisation substantially.}

For our analyses, finding zeros of $D_{I}W$ is efficiently achieved using the root finding methods of \texttt{scipy}~\cite{2020SciPy-NMeth}, especially when compared with homotopy solvers previously employed in~\cite{Martinez-Pedrera:2012teo,Cicoli:2013cha}. From the methods implemented within \texttt{scipy.optimize.root}, we determine via direct comparisons that \texttt{method='hybr'} associated with a modified version of Powell's method~\cite{powell1964efficient} works most reliably and efficiently. As input, we provide a choice of integer fluxes together with initial guesses for the roots. We comment on the sampling of both in the subsequent section. In practice, we run the optimisation module on several CPUs in parallel to speed up the computation. As numerical tolerance $\epsilon_{\text{tol}}$ for the root finding, we use $\epsilon_{\text{tol}}=10^{-10}$ across all models which was chosen as a robust choice when comparing with known special solutions for the torus~\cite{DeWolfe:2004ns} and $\mathbb{P}[1,1,1,6,9]$~\cite{Gallego:2017dvd,Demirtas:2019sip}.

Beyond \texttt{scipy.optimize.root} there are several optimisation methods which could be used. At this stage, however, involving only a small number of checks, we did not find a more efficient method when comparing with e.g.~gradient descent approaches. Clearly, choosing Eqs.~\eqref{eqn:fflat1}, \eqref{eqn:fflat2} as optimisation targets is only one possibility and our framework can be easily applied to the gradient of the scalar potential instead. This is possible due to the efficient automatic differentiation capabilities of our implementation. In the long term, it would be beneficial to apply even more efficient optimisers which can be parallelised easily within the \texttt{JAX} framework and make use of the GPU. These modifications will be the target of the next round of improvements of our framework.

\subsection{Sampling module -- sampling of fluxes and initial guesses}
\label{sec:samplingbias}

In this section, we detail the strategies to efficiently sample flux choices and starting conditions for our root finding algorithm. As we will show subsequently, this vastly outperforms previous access to solutions. Hereby, it is absolutely crucial to employ \texttt{vmap} to guarantee speedups of several orders of magnitude similar to Fig.~\ref{fig:timing_jit}.

Obviously, there remain biases associated to our selection procedure. Although we believe that one should account for these biases using existing statistical techniques, this analysis goes beyond the current scope of the paper.\footnote{An example of such algorithmic biases was discussed recently in~\cite{Krippendorf:2022gcl}.} Here, our main focus is to establish novel methods to numerically access these vacua. The various sampling techniques to be introduced below will be compared in the subsequent section, see in particular Fig.~\ref{fig:CP11169:sampling}.

\subsubsection*{Initial guesses}

Initially, we need to specify points in moduli space from which to start the root finding. To this end, we define a region, e.g.~sphere, box, or the Kähler cone $\cK_{\tilde{X}_3}$, in which points are being uniformly sampled. We treat the size of these regions as a hyperparameter. While the notion of the Kähler cone is mainly useful for LCS limits, conifold regions can be described by spheres centered around the origin.

For our considerations, it is useful to work inside subcones of $\cK_{\tilde{X}_3}$ defined in \eqref{eq:DefKahlerCone}, namely so-called \emph{stretched} Kähler cones defined for some $c\in\bR_{+}$ via~\cite{Long:2016jvd,Demirtas:2018akl}
\begin{equation}\label{eq:DefStrechtedKahlerCone}
    \cK_{\tilde{X}_3}[\, c\,] = \lbrace J\in H^{1,1}(\tilde{X}_3,\bR):\, \text{Vol}_{J}(U)\geq c\; \forall \text{ sub-varieties }U\rbrace\, .
\end{equation}
The tip of $\cK_{\tilde{X}_3}[\, c\,]$ is given by the shortest (in the Euclidean metric) vector $v_{\text{tip}}$ in the Kähler cone $\cK_{\tilde{X}_3}$ for which the generators of the Mori cone $\cM(\tilde{X}_{3})$ (as the dual of its closure) have volume at least $c$. It is convenient to work with $\cK_{\tilde{X}_3}[\, c\,]$ for some $c>0$ because the convergence of $F_{\rm inst}$ is more easily guaranteed.\footnote{To be more precise, the convergence of $F_{\rm inst}$ demands that we stay far enough away from walls of $\cK_{\tilde{X}_3}$ where some of the (mirror) effective curve volumes become zero and the sum in \eqref{eq:InstCorrections} potentially diverges. Clearly, this depends on the structure of (non-zero) GV invariants. For instance, in particular examples, we can analytically continue the LCS periods to regions near conifold singularities by shrinking GV-nilpotent curves~\cite{Demirtas:2020ffz,Alvarez-Garcia:2020pxd}.}
In addition, by picking initial guesses far inside $\cK_{\tilde{X}_3}[\, c\,]$, we are more likely to find roots inside the domain of validity of the LCS expansion, thereby improving the stability and success rate of our root finding methods.

Lastly, we implemented a variant of the cone sampling procedure which is useful in models where the generators of $\cK_{\tilde{X}_3}$ are not known explicitly. In such cases, one can select initial guesses along the one-dimensional subspace $\lbrace c\cdot v_{\text{tip}}:\; c\in\bR_+\rbrace$ (with $v_{\text{tip}}$ here and subsequently denoting the tip of $\cK_{\tilde{X}_3}[\, c=1\,]$) corresponding to the tips of the stretched cones for varying $c$.

\subsubsection*{Flux choices}

We have to specify flux choices where we distinguish the following two approaches:
\begin{enumerate}
    \item {\bf We sample fluxes independently from the starting point:} We do this randomly by sampling from the uniform distribution such that $N_{\text{flux}}$ is below or at a fixed tadpole $Q_{D3}$ as set by Eq.~\eqref{eq:MaxD3Charge}. 
    Similarly, we sample starting points independently from our flux choices using one of the aforementioned techniques. 
    \item {\bf ISD biased sampling:} We introduce a new sampling technique which ensures that both the flux is below the tadpole and the starting point is close to the point satisfying the ISD condition in Eq.~\eqref{eq:ISD_cond_flux}. The idea is to sample only half of the fluxes together with points $(Z_{0}^{i},\tau_{0})$ inside the Kähler cone. Then, the ISD constraint \eqref{eq:ISD_cond_flux} can be solved for the remaining half of the fluxes. We implemented two variants, namely
    \begin{align}
    \label{eq:ISDP} \text{ISD}_{+}&:\quad \tilde{m}_{J}-\tau_{0}\, \tilde{n}_{J} \equiv (\overline{\cN}_{0})_{JI}\left (m^{I}-\tau_{0}\, n^{I}\right )\kom  m^{I},n^{I}\in \bZ\, ,\\
    \label{eq:ISDM} \text{ISD}_{-}&:\quad \tilde{m}^{I}-\tau_{0}\, \tilde{n}^{I} \equiv (\overline{\cN}_0)^{IJ}\left (m_{J}-\tau_{0}\, n_{J}\right )\kom  m_{J},n_{J}\in \bZ\,  .
    \end{align}
    Here, $\cN_{0}$ denotes the gauge kinetic matrix \eqref{eq:GaugeKinMatrix} evaluated at $Z_{0}^{i}$.
    We sample the fluxes on the right hand sides of \eqref{eq:ISDP} and \eqref{eq:ISDM}, while solving for the fluxes on the left.\footnote{A special variant of the ISD$_{+}$ strategy was employed in~\cite{Tsagkaris:2022apo}.} In general, the fluxes $\tilde{m},\tilde{n}$ (dropping indices for convenience) obtained in this way are continuous. To get a meaningful flux vacuum consistent with flux quantisation, we have to round the fluxes $\tilde{m},\tilde{n}\raw m,n\in \bZ$. In effect, this implies that we no longer solve \eqref{eq:ISDP}, \eqref{eq:ISDM} exactly at our initial point in moduli space, but pay the price of shifting the moduli VEVs $\langle Z^{i}\rangle, \langle \tau\rangle$ slightly away from $Z_{0}^{i},\tau_0$, i.e.,
    \begin{equation}\label{eq:shiftISDsampling}
        \langle Z^{i}\rangle = Z_{0}^{i}+\delta Z^{i}\kom \langle \tau\rangle = \tau_{0}+\delta \tau\, .
    \end{equation}
    The sizes of the induced shifts $\lbrace \delta Z^{i}, \delta \tau\rbrace$ from rounding the fluxes depend crucially on the chosen method ISD$_{\pm}$, see Fig.~\ref{fig:CP11169:sampling}. More importantly, the two sampling procedures give rise to different characteristics of the resulting distribution of ISD solutions as we will show in Sect.~\ref{sec:CP11169}.
\end{enumerate}

\subsection{Filter module -- identification of trustworthy string vacua}
\label{sec:consistentvacua}

After computing the stationary points of $V_{F}$ by solving $D_{I}W=0$ or $\partial_{I} V_F=0$, we feed them into a pipeline to extract trustworthy minima.\footnote{It is certainly interesting to analyse the set of all stationary points and compare against results from random matrix theory such as~\cite{Marsh:2011aa,Bachlechner:2012at,Marsh:2015zoa}. We will come back to these questions in the future.} Currently, we implemented the following checks on our solutions:
\begin{itemize}
\item We ensure \emph{positive string couplings and inequivalence to other solutions under $\text{SL}(2,\mathbb{Z})$ transformations}. This is simply because we are interested only in physically relevant solutions with $g_{s}>0$, while avoiding overcounting solutions due to $\text{SL}(2,\mathbb{Z})$. Where necessary, we map $\tau$ to the fundamental domain of $\text{SL}(2,\mathbb{Z})$ by applying translations and S-duality transformations.
\item We verify that our solutions meet the \emph{Kähler cone conditions}. Depending on the available information associated with the background geometry, we use different implementations to verify that the Kähler cone conditions are satisfied. If hyperplanes or generators of the cone are available, then we can easily check whether the given VEVs lie inside the Kähler cone. Alternatively, we check that the Kähler metric is positive definite which is however not a sufficient condition.
\item We check that the \emph{Hessian of the scalar potential is positive semi-definite} by computing its Cholesky decomposition as we find this method to be more efficient than computing eigenvalues directly. This guarantees that the obtained stationary points are actual minima of the scalar potential. In addition, even if non-negativity of the Hessian can be guaranteed, ensuring the \emph{absence of flat directions} requires special attention. This is because it can be numerically hard to distinguish minima from flat direction, especially when large hierarchies of scales are involved. For vacua with $D_{I} W=0$, the number of massless fluctuations around the minimum, i.e., the dimension of the remaining moduli space is given by the rank of the matrix\footnote{The diagonal entries of this matrix (evaluated at a minimum) are proportional to the mass matrix of chiral fermions.} (see e.g.~\cite{deAlwis:2013jaa})
\begin{equation}
M\bigl |_{D_{I}W=0}=\left (\begin{array}{cc}
D_{I}D_{J}W & K_{I\ov J}W \\ 
K_{\ov I  J}\ov W & D_{\ov I}{D_{\ov J}\ov W}
\end{array} \right )\, .
\end{equation}
For vacua with $D_{I}W\neq 0$, we have to consider the number of first order obstructions to $\p_{I}V_F=0$ which is simply the rank of the Hessian. In both cases, the rank is computed using the tolerance $\epsilon_{\text{tol}}$ used in the optimisation module.
\item \emph{Validity of the LCS approximation}: We check Eq.~\eqref{eq:InstCutoff} for $\varepsilon=0.1$.
This gives a rough estimate on the radius of convergence for degrees available in each model.\footnote{Arguably, to fully trust a solution, one would need to compute the radius of convergence to high precision. We intend to discuss this problem in a separate work.}
\end{itemize}
A solution passing these criteria we refer to as a flux solution or vacuum. All of the above checks are fully \texttt{vmap} compatible (recall Sect.~\ref{sec:jax}).

\section{Numerical experiments}
\label{sec:numericalexperiments}

Having described our algorithm, we now showcase some initial capabilities and comment on physical implications of the respective results. To begin -- using the well-studied example of a degree $18$ hypersurface in the weighted projective space $\bP[1,1,1,6,9]$ -- we discuss the differences in our sampling procedures. We then generate large samples of vacua for four and five moduli models which were introduced in~\cite{Cicoli:2013cha} featuring at least one million flux vacua, vastly outperforming the handful of previously obtained vacua. Interestingly, we identify that the respective $|W_0|-$distributions can be well approximated by the same probability distribution. Finally we analyse the scaling behaviour with $h^{1,2}$ by studying examples with up to $25$ moduli; identifying suitable success rates and timing behaviour of our approach.

\subsection{Sampling procedures in practice -- $h^{1,2}=2$}\label{sec:CP11169}

To characterise the physics associated with the distribution of flux vacua, we need to understand the biases arising from the respective sampling techniques.\footnote{Note that this applies to both numerically and analytically constructed flux vacua.} To start the discussion, we present the difference of the vacua distribution in our sampling techniques previously introduced in Section~\ref{sec:samplingbias}. To allow for simpler visualisation, we study the degree 18 hypersurface $X_3$ in weighted projective space $\bP[1,1,1,6,9]$~\cite{Candelas:1994hw} with Hodge numbers $(h^{1,1},h^{1,2})=(2,272)$.

\begin{table}[t!]
\centering
\begin{tabular}{|c||c|c|c|c|c|}
\hline 
\diagbox[width=3.4em,height=2em]{$d_{1}$}{ $d_{2}$}& 0 & 1 & 2 & 3 & 4 \\
\hline 
\hline 
 &  &  &  &  &   \\[-1.1em]
0 & 0 & 3 & -6 & 27& -192  \\[0.2em]
\hline 
 &  &  &  &  &   \\[-1.1em]
1 & 540 & -1,080 & 2,700 & -17,280 & 154,440  \\[0.2em]
\hline 
 &  &  &  &  &   \\[-1.1em]
2 & 540 & 143,370 & -574,560 & 5,051,970 & -57,879,900  \\[0.2em]
\hline 
 &  &  &  &  &   \\[-1.1em]
3 & 540 & 204,071,184 & 74,810,520 & -913,383,000 & 13,593,850,920  \\[0.2em]
\hline 
 &  &  &  &  &   \\[-1.1em]
4 & 540 & 21,772,947,555 & -49,933,059,660 & 224,108,858,700& -2,953,943,334,360  \\[0.2em]
\hline 
\end{tabular} 
\caption{The leading order GV invariants for $\bP[1,1,1,6,9]$.}\label{tab:GVsCP11169} 
\end{table}

In detail, we focus on a particular locus in moduli space where the CY is invariant under $G=\bZ_{6}\times \bZ_{18}$~\cite{Giryavets:2003vd}. By restricting to fluxes invariant under this symmetry, we only need to solve the $F$-term conditions along the invariant subspace. The corresponding periods can be obtained from the mirror CY giving rise to an effective prepotential depending only on two moduli. The associated topological data is
\begin{align}
\kappa_{111}&=9\kom \kappa_{112}=3\kom\kappa_{122}=1\kom a_{ij}=\dfrac{1}{2}\left (\begin{array}{cc}
9 & \;\;3 \\ 
3 & \;\;0
\end{array} \right )\kom b_{i}=\dfrac{1}{4}\left (\begin{array}{c}
17 \\ 
6
\end{array} \right )\, .
\end{align}
We also computed GV invariants up to degree 100 using methods described in~\cite{Demirtas:2023als} and collected the leading order GV invariants (see Tab.~\ref{tab:GVsCP11169} as a reference). We consider an orientifold with $h^{1,2}_{+}=0$ constructed explicitly in~\cite{Louis:2012nb} for which the D3-tadpole is given by $Q_{D3}=276$.

The Kähler cone $\cK_{\tilde{X}_3}$ is simply the positive quadrant generated by the vectors $\lbrace (1,0),(0,1)\rbrace$. One thus easily samples points inside the cone by taking positive linear combinations of these generators. The tip of the stretched cone $\cK_{\tilde{X}_3}[c=1]$ is just given by $v_{\text{tip}}=(1,1)$, recall Sect.~\ref{sec:samplingbias}.

\subsubsection*{Benchmarking our performance}

Initially, we compare our implementation with existing searches using the Paramotopy method~\cite{2018arXiv180404183B} which has been employed for our model in~\cite{Martinez-Pedrera:2012teo} to search for flux vacua with $N_{\text{flux}}=34$. While the scan of~\cite{Martinez-Pedrera:2012teo} was performed only including classical terms, we are able to conduct two separate searches: the first uses the purely classical prepotential, whereas the second includes instanton terms up to degree 10. We note that in our framework there is no significant change in performance for either of the two scans despite the fact that $F_{\text{inst}}$ includes 65 additional terms with non-vanishing GVs.

Our calculations have been performed on the LMU cluster on a single node with 4 cores with 5GB of memory and 50,000 input fluxes. For the two runs, we found 34,542 (with instantons) and 33,019 (classical) vacua with $\varepsilon=0.1$ in \eqref{eq:InstCutoff} in about 45 minutes respectively.  From those vacua, using sampling method ISD$_{+}$, 33,047 and 31,042 survive when choosing instead $\varepsilon=0.01$ in \eqref{eq:InstCutoff} implying that the LCS expansion is well under control in our framework. As a comparison, the equivalent scan in~\cite{Martinez-Pedrera:2012teo} was performed with 100 nodes each with 32 cores and required a total calculation time of around 75,000 hours to find around 24,882 acceptable solutions.

\subsubsection*{Qualitative comparison of sampling approaches}

\begin{figure}[t!]
    \centering
    \includegraphics[width=1.\textwidth]{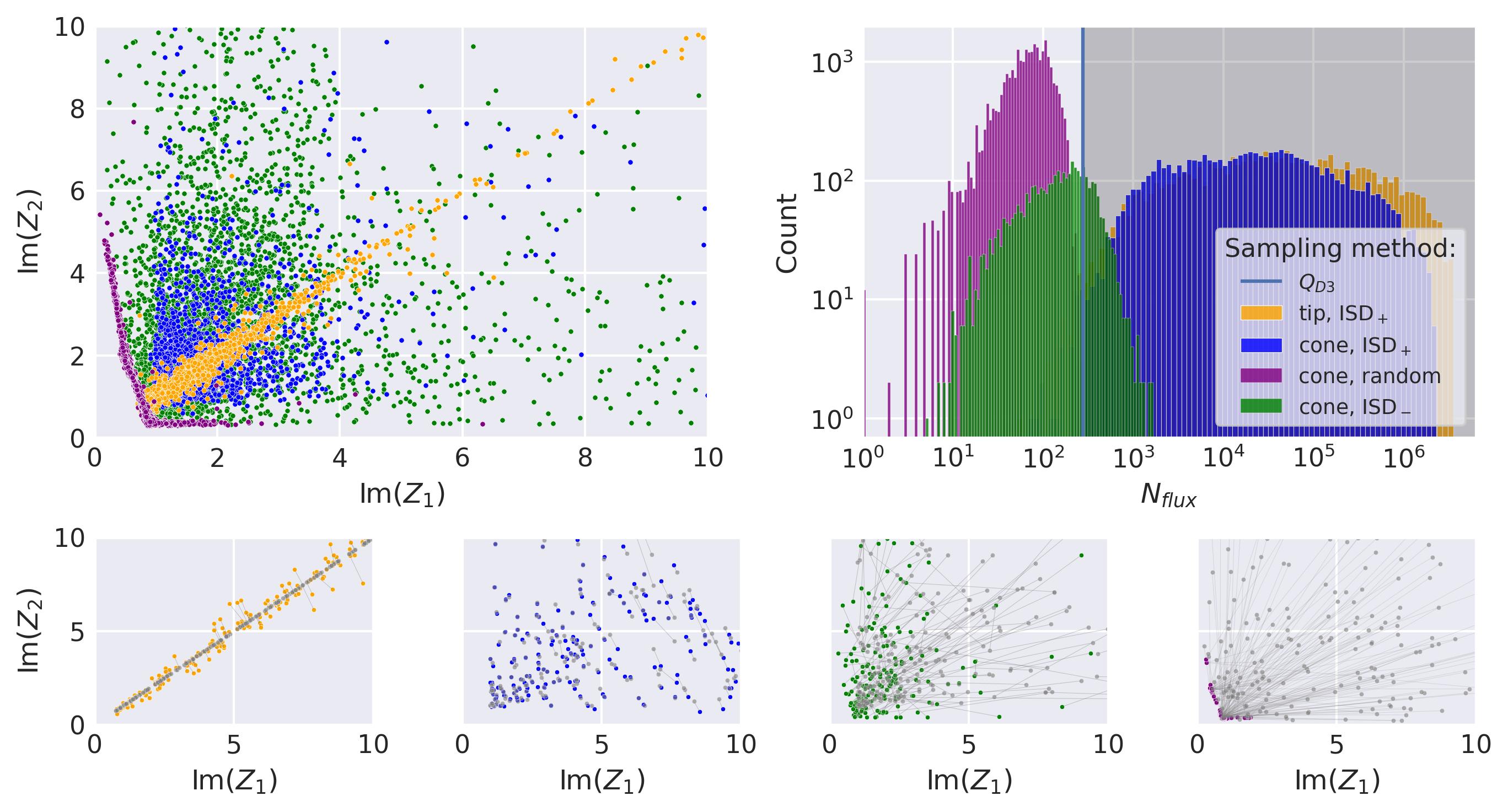}
    \caption{{\bf Comparison between the different sampling procedures.} \emph{Top left}: Distribution of flux vacua in the plane of moduli VEVs, utilising units as outlined in the main text. \emph{Top right:} Distribution of $N_{\text{flux}}$ with the vertical line indicating the tadpole bound $Q_{D3}=276$ and the dark shaded region highlighting solutions with $N_{\text{flux}}>Q_{D3}$. \emph{Bottom row:} Distribution of initial guesses (gray) compared to the actual solutions to $D_I W=0$.  }
    \label{fig:CP11169:sampling}
\end{figure}

Turning to the differences of our sampling methods, we show the distribution of moduli VEVs for the different sampling methods in Fig.~\ref{fig:CP11169:sampling}. By sampling points only along tips of the stretched Kähler cone for different choices of $c$ (recall \eqref{eq:DefStrechtedKahlerCone}), the resulting sample of flux vacua remains largely 1-dimensional (orange). For cone sampling, one covers a much broader range inside the Kähler cone for both ISD$_{\pm}$. Interestingly, randomly sampling fluxes results in moduli VEVs clustered at the boundary of the Kähler cone. Being close to the boundary suggests that these solutions may not survive in the presence of higher order instanton corrections as the region with $|F_{\rm inst}/F|<\varepsilon$ shifts further inside the Kähler cone when cutting the instanton prepotential at higher degree.

Looking at the right in Fig.~\ref{fig:CP11169:sampling}, we find that the distribution of flux tadpole contributions $N_{\text{flux}}$ are vastly different among the sampling methods. We clearly see that the $\text{ISD}_{+}$ method leads to rather large tadpoles, even for models with small $h^{1,2}$.\footnote{This is reminiscent of the observations made in~\cite{Tsagkaris:2022apo} for a model with rather large $h^{1,2}=50$.} In contrast, the $\text{ISD}_-$ approach generates samples with significantly smaller $N_{\rm flux}$ values. 

The bottom row of Fig.~\ref{fig:CP11169:sampling} shows a subset of the moduli VEVs at the minimum and the associated initial guesses. The relative distance between them depends significantly on the sampling method. These noticeable dissimilarities can lead to algorithmic biases (see~\cite{Krippendorf:2022gcl}) as the identification of minima is subject to input parameters of the optimisation module like the maximal number of steps. For random flux sampling, we find the largest displacements where, despite sampling points far inside the cone, the moduli VEVs end up close to the boundary. In contrast, the induced shifts $\delta Z^i$ (recall Eq.~\eqref{eq:shiftISDsampling}) from rounding fluxes in ISD sampling are appreciably smaller, though they are larger for ISD$_{-}$ than for ISD$_{+}$. This is because, when using the ISD$_{-}$ method, the inverse matrix $\overline{\cN}^{IJ}$ appears in \eqref{eq:ISDM}. In effect, this amplifies the error in \eqref{eq:ISDM} when rounding the continuous fluxes on the left hand side. Hence, the actual solutions to the $F$-flatness conditions are on average further away from the initial guess than in ISD$_{+}$, thereby also affecting the success rate.

To conclude, we clearly see that the different biases due to the sampling method have indeed an appreciable effect. These biases have to be accounted for to deduce constraints on the space of flux vacua and to extract probabilistic statements about the string landscape.\footnote{We note that dealing with such biases in practice is ubiquitous in observational cosmology, see for instance~\cite{Clerc:2022yan} for examples in the context of large scale structure cosmology constraints.} At this stage we postpone a quantification of these effects to future work as we believe that these techniques are best applied in the context of a direct physics question.

In particular, let us highlight that already in this simple example the distribution of $\NF$ depends critically on the chosen sampling and solving methods. As is obvious from Fig.~\ref{fig:CP11169:sampling}, large values of $\NF\gtrsim \mathcal{O}(10^{3})$ are naturally preferred for ISD$_+$ sampling. This should however not be understood as evidence for the tadpole conjecture. It is merely a sign that, due to the high dilution of string vacua with low $\NF\leq Q_{D3}$ compared to swampland solutions with high $\NF>Q_{D3}$, randomly sampling points in $\cK_{\tilde{X}_3}$ generically leads to large $\NF$. Obviously, this problem becomes even more severe at higher $h^{1,2}$, see also Sect.~\ref{sec:Largeh12}. Hence, it becomes necessary to optimise the procedure for the sampling of points in the Kähler cone.

\subsection{Generating samples of flux vacua -- $h^{1,2}=4,5$}

Capabilities to generate vast sets of generic flux vacua in regimes with $h^{1,2}>2$ has so far been rather limited. Here, we would like to demonstrate the ease with which such datasets can be efficiently produced within our framework even for more elaborate geometries. We therefore revisit models with four and five complex structure moduli presented in~\cite{Cicoli:2013cha}.

Specifically, we study three CY hypersurfaces presented in the appendix\footnote{One easily verifies from the list of orientifolds in~\cite{Crino:2022zjk} that the model discussed in the main text of~\cite{Cicoli:2013cha} does not allow for an orientifold with $h^{1,2}_{+}=0$.} of~\cite{Cicoli:2013cha}, though we construct orientifolds from different $\mathbb{Z}_2$-involutions.\footnote{The orientifold examples of~\cite{Cicoli:2013cha} were specifically constructed to satisfy certain model building criteria presented in~\cite{Cicoli:2012vw} for D-branes at singularities with $h^{1,1}_{-}\neq 0$.} The latter are obtained from the orientifold database of~\cite{Crino:2022zjk} with the D3-tadpole values $Q_{D3}=104$ and $Q_{D3}=192$ for $(h^{1,1},h^{1,2})=(4,98)$ (example 2) and $(h^{1,1},h^{1,2})=(5,185)$ (examples 3, 4) respectively.

\begin{figure}[t!]
    \centering
    \includegraphics[width=0.49\textwidth]{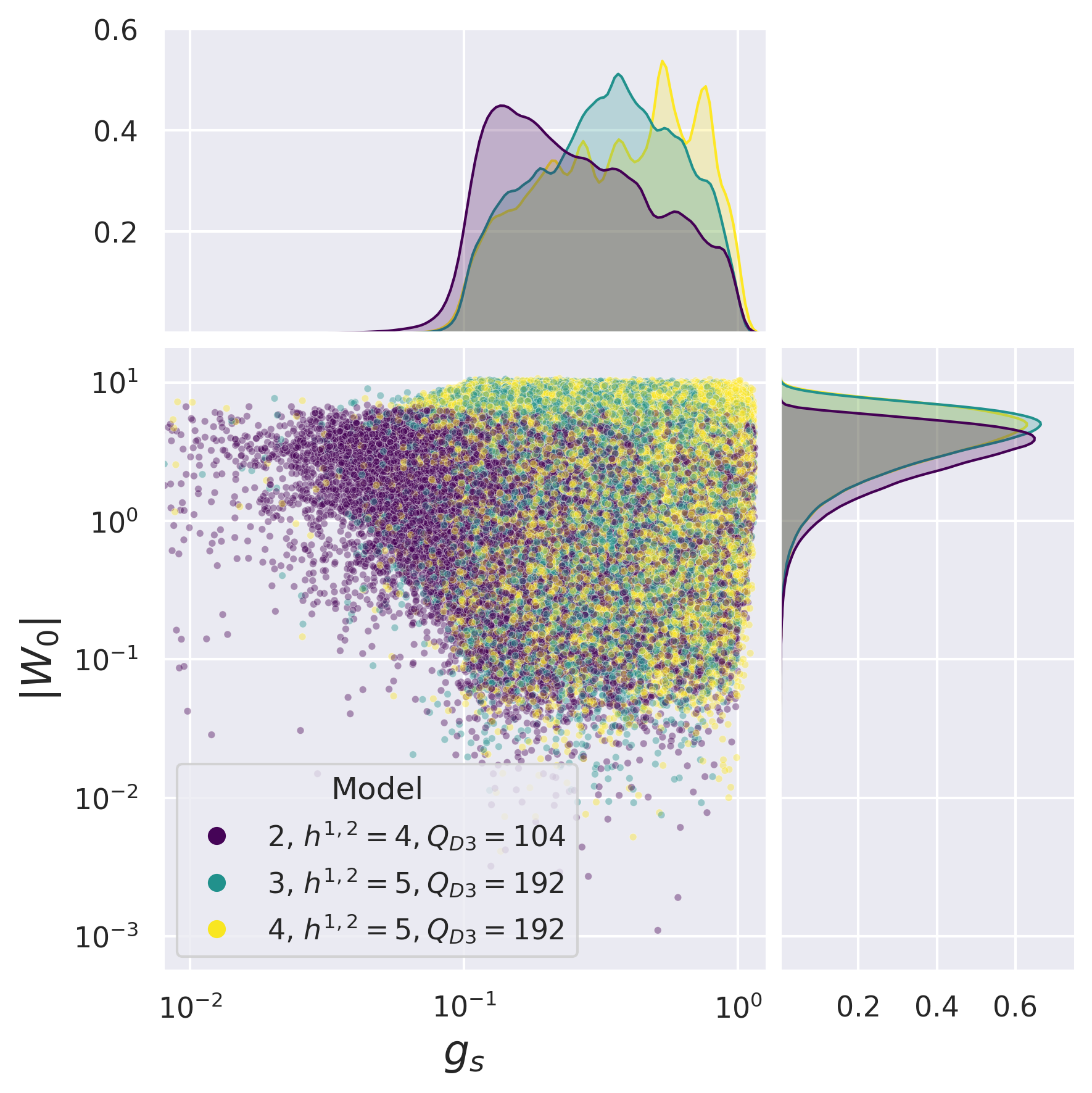}
    \includegraphics[width=0.49\textwidth]{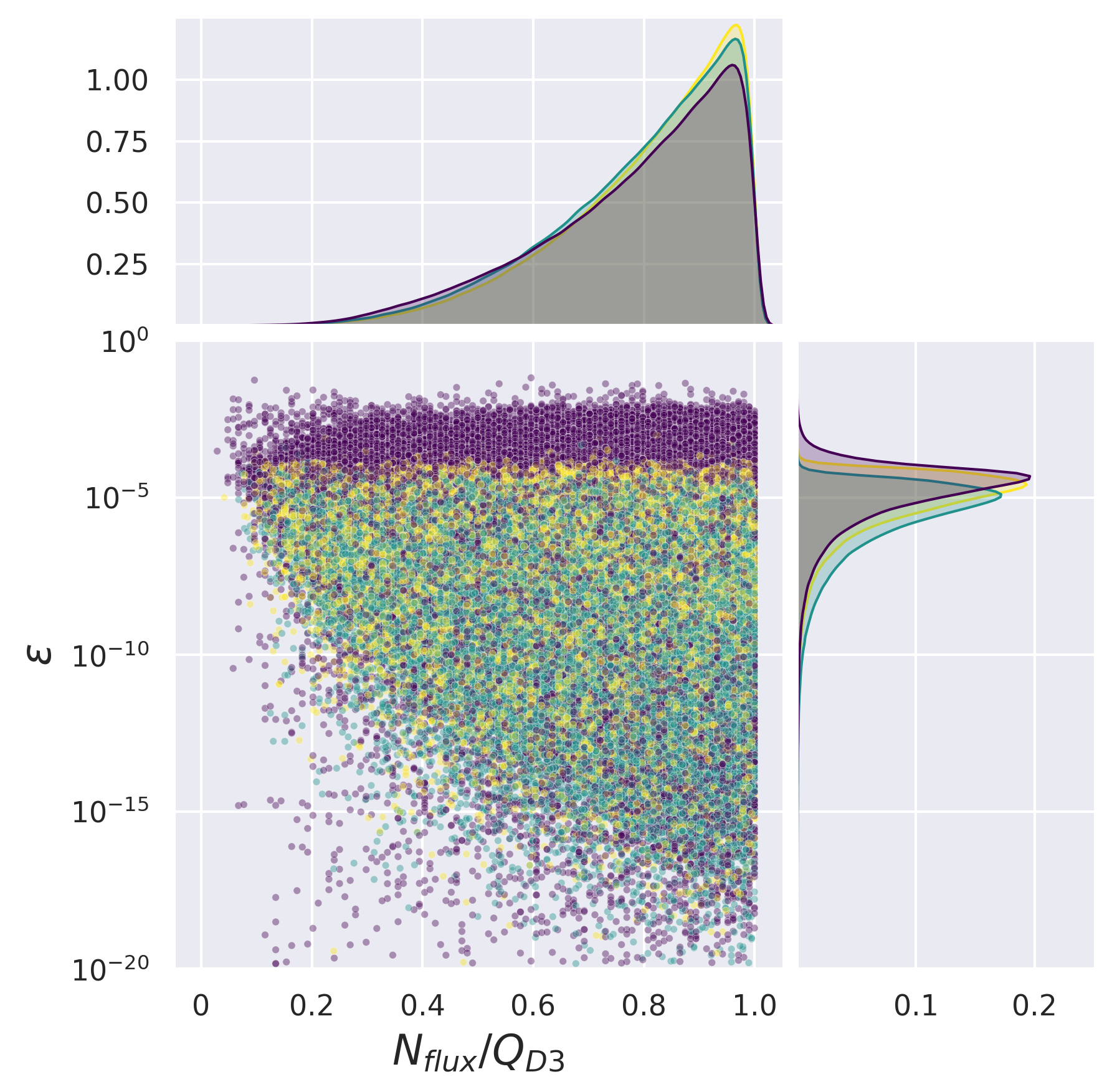}
    \caption{Samples of flux vacua with four and five complex structure moduli of~\cite{Cicoli:2013cha}. \emph{Left:} Distribution of $W_{0}$ and $g_{s}$, showing universal behaviour in the distribution of $W_0$. \emph{Right:} Distribution of $\varepsilon$ and $N_{\text{flux}},$ demonstrating very good control over the instanton expansion.}
    \label{fig:1312W0gs}
\end{figure}

As in the previous example, we study these models near a \emph{Greene-Plesser point}~\cite{Greene:1990ud} where the CY is manifestly invariant under a discrete subgroup $G$. Turning on fluxes only along the invariant 3-cycles allows to fix the non-invariant moduli fields at their fixed point under $G$, thereby solving the associated $F$-flatness conditions. The remaining periods can directly be computed on the mirror dual CY $\tilde{X}_3$ and depend only on the invariant moduli of which we have four (example 2) or five (examples 3,4) in our models.

For these example, we now collect flux vacua by using the ISD$_{+}$ sampling method for the fluxes and initial points. To this end, we randomly generate fluxes in the range $[-5,5]$ and points inside the Kähler cone up to Euclidean distance $10$ from the origin using the cone generators. For each model, we collected approximately $10^6$ vacua taking into account instanton corrections up to degree $10$ which took 10 hours of run time on a machine with 4 CPUs and 10GB memory.

We show the distribution of $|W_{0}|$ and $g_{s}$ on the left of Fig.~\ref{fig:1312W0gs}. Apart from a slight shift of the peak, the former seems to be almost identical across the three models. This is a hint at universality of the $|W_{0}|$ distribution across different models within our current framework. In contrast, the distribution of $g_{s}$ exhibits non-trivial structures which are distinct for each model. Clearly, at this point, we are unable to infer that these are characteristic features of the respective models given the expected biases in our data, recall Sect.~\ref{sec:samplingbias}. Nonetheless, it is interesting to point out that, while the distribution for example 2 peaks at small $g_s$, we observe two distinctive peaks at large $g_s$ for example 4.

The distribution of $\varepsilon$ and $N_{\text{flux}}/Q_{D3}$ is displayed on the right of Fig.~\ref{fig:1312W0gs}.
While in the analysis of~\cite{Cicoli:2013cha} most vacua were unstable against the inclusion of higher order instanton corrections, we find overall an excellent control over the instanton expansion as exhibited by small $\varepsilon$. This can be expected given that our sampling methods allow us to stabilise moduli deep inside the Kähler cone where instanton terms are highly suppressed. In contrast, as we have seen previously in Fig.~\ref{fig:CP11169:sampling}, random sampling fluxes as in~\cite{Cicoli:2013cha} tends to drive the $F$-term solutions closer to the boundary, thereby loosing control over the instanton expansion. This makes evident the fact that our new sampling techniques are superior to previously employed techniques.

The rescaled distribution of $N_{\text{flux}}/Q_{D3}$ is largely model independent. At first sight, this is surprising given the expected scaling \eqref{eq:DenefDouglasNumVacua} of the number of vacua with $h^{1,2}$ and the tadpole.
We believe that this is due to the chosen sampling method, though this observation deserves further scrutiny.

\subsection{Scaling behaviour for large numbers of moduli -- $h^{1,2}\leq 25$}\label{sec:Largeh12}

Having seen the capabilities to generate reliable samples, we next comment on the capabilities of this framework to analyse geometries with even larger number of moduli. To this end, we construct examples with $h^{1,2}\in \lbrace 5, 10, 15, 20,25\rbrace$ moduli using \texttt{CYTools}, see Tab.~\ref{tab:largeh12models} and ancillary files for the details regarding these models. Specifically, we selected trilayer polytopes~\cite{Moritz:2023jdb} in KS with the largest $h^{1,1}$ for given $h^{1,2}$ and constructed smooth $O3/O7$ orientifolds with $h^{1,1}_{-}=h^{1,2}_{+}=0$. Even though we collected these models by requiring, in addition to the existence of suitable orientifolds, large $Q_{D3}$, we note that similar analysis could be done with much smaller $Q_{D3}$ at the prize of a higher run time.

Given the model data, we look for flux vacua with $N_{\rm flux}\leq Q_{\rm D3}$. We sample vacua by using the ISD$_+$ method running on machines with 8 CPU cores and 12GB RAM. For each model, we include instanton contributions up to degree $2$.

\begin{table}[t!]
    \centering
    \begin{tabular}{c|c|c|c|c|c}
         $h^{1,1}$ & $h^{1,2}$ & $Q_{D3}$ & success rate & $\sharp$vacua & $\text{min}(N_{\text{flux}})$ \\
         \hline
         \hline
         213 & 5 & 220 & $50\%$ & 1,370,842 &  5\\
         \hline
         244 & 10 & 256 & $16\%$ & 498,545 & 36 \\
         \hline
         399 & 15 & 416 & $7\%$ & 168,291 & 116 \\
         \hline
         350 & 20 & 372 & $<1\%$ & 36 & 180\\
         \hline
         245 & 25 & 272 & $<1\%$ & 1 & 270\\
    \end{tabular}
    \caption{Hodge numbers and tadpole values for the selected models.}
    \label{tab:largeh12models}
\end{table}

The average success rates of finding vacua from some input fluxes is listed in Tab.~\ref{tab:largeh12models} where we notice a significant drop at $h^{1,2}\geq 20$. We believe that this is because of several constraining factors conspiring to drastically decrease the success rate at large $h^{1,2}$. Firstly, high dimensionality of flux and moduli spaces from which we sample typically means slower evaluation time. In addition, it becomes even harder to perform numerical optimisation, i.e.,~to find solutions to $D_I W=0$. Lastly, a single phase of the Kähler cone becomes narrower~\cite{Demirtas:2018akl}, while the number of phases of the extended Kähler cone increases exponentially. That is, we see less and less of the complete moduli space. This becomes e.g.~relevant in ISD sampling where the induced shifts in \eqref{eq:shiftISDsampling} may easily lead to moduli VEVs $\langle Z^i\rangle, \langle\tau\rangle$ outside the original Kähler cone. All in all, the chances get rather slim to locate minima at large $h^{1,2}$ below a certain threshold of $Q_{D3}$. Clearly, despite significant progress, we point out that our methods are simply not refined enough yet to access these regimes. 

In this context, we stress that sampling below a realistic value of $Q_{D3}$ is many orders of magnitude harder than sampling with arbitrary $N_{\text{flux}}$ by virtue of \eqref{eq:DenefDouglasNumVacua}. We tested our algorithm on examples with $h^{1,2}>100$ and easily obtained solutions with $N_{\text{flux}}\gg Q_{D3}$. This is just telling us that sampling from an effectively continuous distribution for swampland solutions is much easier than probing a discrete distribution of landscape solutions.

\begin{figure}[t!]
    \centering
    \includegraphics[width=0.99\textwidth]{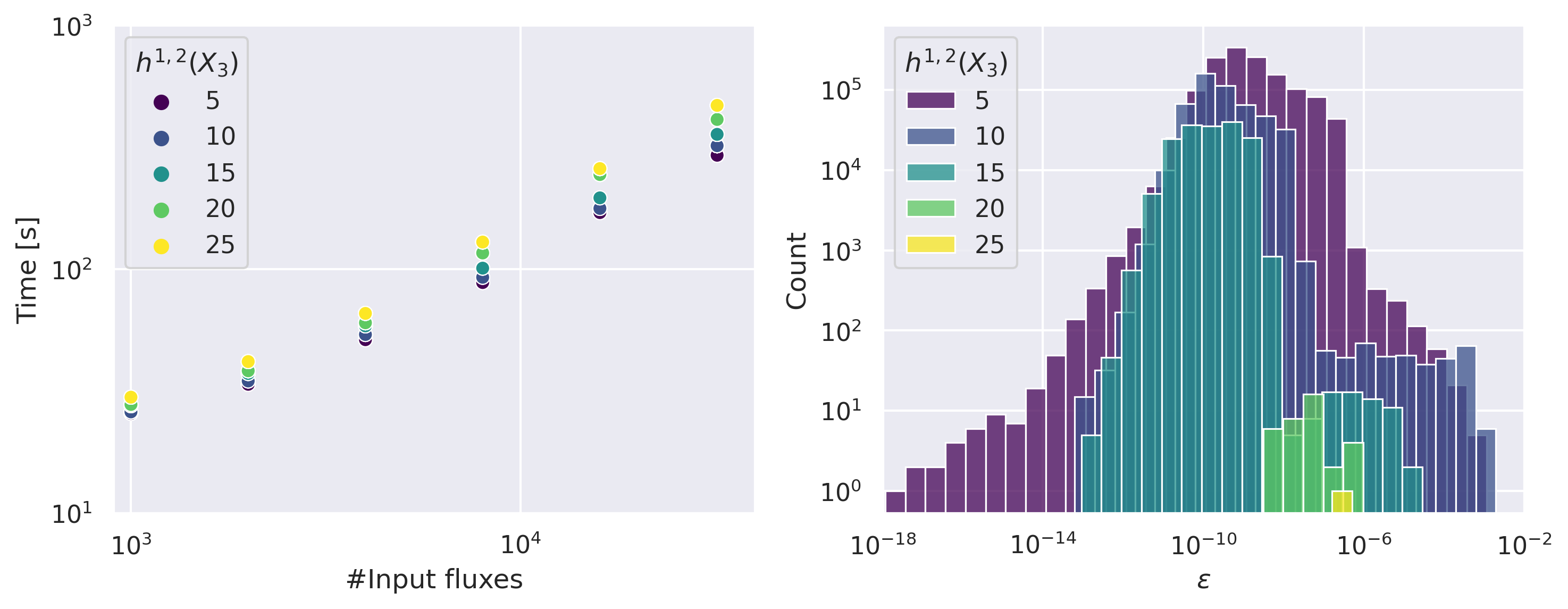}
    \caption{
    Efficiency analysis of our code for the models summarised in Tab.~\ref{tab:largeh12models}: \emph{Left:} Time required to apply the optimisation module on a given number of input fluxes obtained from the average over five independent runs. \emph{Right:} The distribution of the instanton control parameter $\varepsilon$ (recall Eq.~\eqref{eq:InstCutoff}). Note that all $\varepsilon$ values are far below our cut-off $0.1.$
    }
    \label{fig:timing_root_finding}
\end{figure}

We present the performance of our code in Fig.~\ref{fig:timing_root_finding}. On the left, we show the time required to obtain roots to the $F$-term conditions by employing our optimisation module. Crucially, we observe virtually no scaling with respect to $h^{1,2}$ which, as we said before, is largely due to \texttt{jit}-compilation.\footnote{Recall that \texttt{vmap} is not being used currently in our optimisation module which would lead to a significant speed up over a few orders of magnitude. We however use \texttt{vmap} for all other parts of the algorithm e.g.~when checking the validity of our solutions.} This in principle opens up the opportunity to efficiently pursue systematic scans for flux vacua over a large range of models within the wider CY landscape.

On the right in Fig.~\ref{fig:timing_root_finding}, we plot the distributions of $\varepsilon$ computed for instanton contributions up to degree $2$ via Eq.~\eqref{eq:InstCutoff}. We again find that our solutions have extremely small $\varepsilon \lesssim 10^{-2}$ signifying excellent control over instanton corrections. Nonetheless, as we stressed several times before, a more thorough analysis is a prerequisite to make definitive statements about the radius of convergence of the sum over instanton terms.

\begin{figure}[t!]
    \centering
    \includegraphics[width=1.\textwidth]{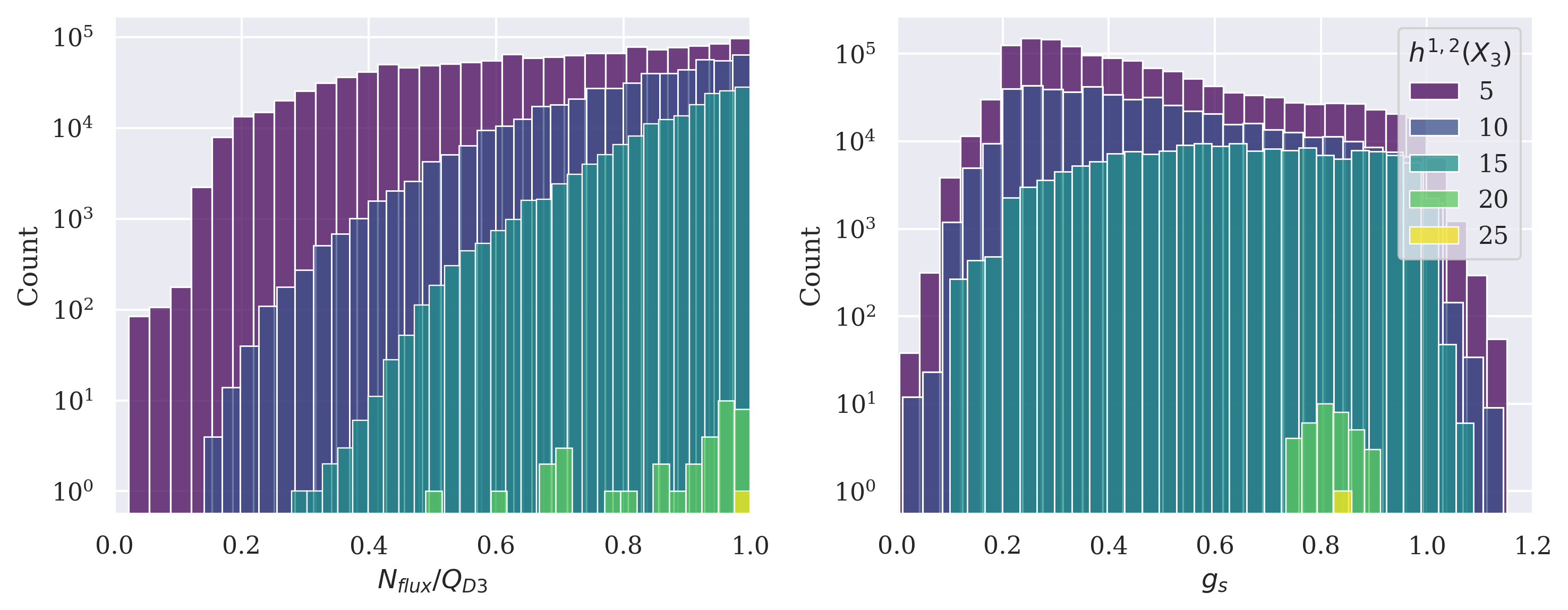}
    \caption{Solutions for large number of moduli. \emph{Left:} Distribution of the ratio of $N_{\text{flux}}$ over the maximally allowed D3-charge $Q_{\text{D3}}$. \emph{Right:} Distribution of the string coupling $g_s$.} 
    \label{fig:Nflux_Largeh12}
\end{figure}

The left plot in Fig.~\ref{fig:Nflux_Largeh12} shows the distribution of the ratio of $N_{\text{flux}}$ over the maximally allowed D3-charge $Q_{\text{D3}}$. (Note that each sample at fixed $h^{1,2}$ contains a different number of vacua as summarised in Tab~\ref{tab:largeh12models}.) As we argued before in Sect.~\ref{sec:CP11169}, our ISD$_+$ sampling, which we picked because of its reliability, prefers large $N_{\text{flux}}$ which is why the distribution in Fig.~\ref{fig:Nflux_Largeh12} is peaked near $N_{\text{flux}}/Q_{\text{D3}}\sim 1$. In contrast, $g_s$ is almost uniformly distributed in a wide range of values for $h^{1,2}\leq 15$, though the distribution for $h^{1,2}=5,10$ exhibit moderate peaks around $g_s\approx 0.2$.

\section{Conclusions}
\label{sec:Con}

We presented an efficient and reliable framework to find flux vacua providing access to largely unexplored territories of string theory solutions. These advances have been achieved by combining various numerical techniques ubiquitously used in the context of machine learning, namely automatic differentiation, just-in-time compilation, and vectorisation. These methods allow for the generation of fast (C++), parallelisable (on CPU and GPU), and versatile code allowing a simple interface with EFT quantities, e.g.,~the couplings in the prepotential.

In this work, we are only hinting at upcoming physics analysis for these flux solutions. For instance, our framework, \texttt{JAXVacua}, intrinsically calculates the scalar potential and its derivatives at machine precision to any order in some approximation scheme. This enables systematic studies of properties of moduli potentials in general classes of models. Even more importantly, as we showed in the main text, large scale surveys of previously inaccessible regimes in the string landscape become feasible. This provides excellent opportunities to refine our understanding of the statistics of vacuum solutions, their associated mass spectra and couplings to visible matter fields. Ultimately, there is a wide range of applications of our methods to string model building and cosmology.

At this stage, there are a few technical improvements which we leave for the near future to make flux vacua at LCS of the entire KS database accessible:
\begin{itemize}
    \item \textbf{Optimisation:} 
    Although we identify \texttt{scipy.optimize.root} as an efficient root finding method for flux vacua, it is nevertheless a limiting factor in efficiency. For instance, it would allow for further speed-up if we can use this method on the GPU directly. In this context it is interesting to see whether other optimisers, such as variants of gradient descent can lead to more efficient flux vacua generation, see~\cite{Comsa:2019rcz} for applications of gradient descent methods to different types of string theory vacua. As the successful choice of optimiser is also based on the structure of the respective energy landscape in this optimisation problem, it will be interesting to compare the similarities between the string theory landscape and other energy landscapes (e.g.~spin glass systems or deep learning optimisation).
    \item \textbf{Characterising the trustable EFT regime:} To understand the stability of flux solutions at the boundaries of the LCS expansion, efficient ways of estimating the radius of convergence in our examples are pending (see~\cite{Hosono:1993qy,Hosono:1994ax,Candelas:1994hw,Klemm:1999gm} for examples). This applies to a subset of our solutions where the control parameter ${\cal F}_{\rm inst}/{\cal F}_{\rm pert}=\varepsilon$ becomes large. Currently, we evaluate this parameter by calculating the GV expansion using \texttt{CYTools}~\cite{Demirtas:2023als} up to high degrees. In this way, we identified a large sample of solutions where these contributions are exponentially suppressed up to degree 10. Crucially, our implementation is able to compute the relevant quantities to any order in the GV expansion at machine precision.
    \item \textbf{Sampling input data:} We observed that the success rate of finding vacua below tadpole decreases in our current pipeline significantly at large $h^{1,2}>15$. We believe that this is mainly due to inefficient sampling procedures. To make progress, it would be instructive to formulate conditions that, for a given choice of fluxes, guarantee solutions to exist inside the Kähler cone. Secondly, while the ISD sampling methods have shown outstanding success, it turns out to be hard to sample points in the Kähler cone such that $N_{\text{flux}}\leq Q_{D3}$. By implementing this as an optimisation problem, gradient descent methods could help locating such points inside the Kähler cone.
\end{itemize}

We note that, although our methods and examples are currently limited to LCS limits, our implementation can be adapted to study different asymptotic limits in moduli space such as conifold regimes~\cite{Demirtas:2020ffz,Alvarez-Garcia:2020pxd}. In these cases, the solutions to the Picard-Fuchs equation for the periods have different analytic properties. Such generalisations will be investigated in the future.

More generally speaking, our approach in \texttt{JAXVacua} can be extended to other moduli, such as Kähler moduli stabilisation. This will enable a general search for vacua where the presence of novel minima could be observed that fall neither in the class of the standard KKLT nor the LVS. Such new solutions are believed to arise from the intricate structure of EFTs from string compactifications. In the presence of large numbers of moduli, such regions are most easily discovered in comprehensive numerical explorations as presented in this paper.

\section*{Acknowledgements}

We would like to thank Alex Cole, Mathis Gerdes, Arthur Hebecker, Manki Kim, Severin Lüst, Liam McAllister, Jakob Moritz, Richard Nally and Gary Shiu for useful discussions. We especially thank Andres Rios-Tascon for providing the code to compute GV and GW invariants. AS thanks DAMTP at the University of Cambridge and Ludwig Maximilian University of Munich for hospitality where parts of this work have been completed. The research of AS is supported by NSF grant PHY-2014071.

\bibliographystyle{utphys}
\bibliography{Literatur}

\end{document}